\documentstyle[11pt,aaspp4,flushrt]{article}
\def \beq{\begin{equation}}
\def \beqa{\begin{eqnarray}}
\def \eeq{\end{equation}}
\def \eeqa{\end{eqnarray}}
\def \al{\alpha}
\def \bx{{\bf x}}
\def \by{{\bf y}}
\def \bal{{\bf \alpha}}
\def \sb{\bar{s}}
\def \ub{\bar{u}}
\def \nb{\bar{\nu}}
\def \sign{\rm sign}

\begin{document}

\title{Fast calculation of a family of elliptical mass gravitational lens models}

\author{Rennan Barkana\footnote{email: \verb+barkana@ias.edu+}}
\affil{Institute for Advanced Study, School of Natural Sciences, Princeton, NJ 08540}

\begin{abstract}
Because of their simplicity, axisymmetric mass distributions are often
used to model gravitational lenses. Since galaxies are usually observed
to have elliptical light distributions, mass distributions with 
elliptical density contours offer more general and realistic lens 
models. They are difficult to use, however, since previous studies
have shown that the deflection angle (and magnification) in this
case can only be obtained by rather expensive numerical integrations. 
We present a family of lens models for which the deflection can be 
calculated to high relative accuracy $(10^{-5})$ with a greatly
reduced numerical effort, for small and large ellipticity alike. This
makes it easier to use these distributions for modelling individual
lenses as well as for applications requiring larger computing times,
such as statistical lensing studies. A program implementing this
method can be obtained from the author\footnote{or at 
\verb+http://www.sns.ias.edu/\~{}barkana/ellip.html+}.
\end{abstract}


\vspace{.1in}

\section{Introduction}

The types of density profiles used to model gravitational lenses 
have been motivated by observations of lenses in addition to
practical considerations. Observed features of galaxies and clusters
that can be incorporated into lens models include ellipticity and
radially decreasing mass density profiles. But it is essential to be 
able numerically to calculate quickly the deflection angle and
magnification of a light ray due to a lens model, in order to probe
the entire range of parameter space when searching for a best fit
to an observed lens. Numerical efficiency is even more important
in cases with multiple sources such as radio lenses observed
at high resolution. Strongly lensed arcs in clusters sometimes
lie near cluster galaxies and thus modelling all the lensed 
features in a cluster may require a combination of many individual
lenses. Another application which depends on numerical speed is the
statistical study of properties of the images of a lens model 
for comparison with lens surveys.

In attempting to construct models that are as realistic as possible,
new features must be introduced carefully, since in a single case
of multiple lensing there are a small number of constraints requiring
a small number of model parameters. Ellipticity adds just two 
parameters (magnitude and orientation), and it appears to be essential.
Galaxies and clusters often appear elliptical, with significant
numbers having axis ratios $b/a$ smaller than $0.5$. Axisymmetric
lens models are also excluded by many observed lens systems, in
which the images are not colinear with the lens position on the
sky. Of course, the asymmetry in each case may also be due in part to 
external shear from other nearby galaxies or from large-scale
structure along the line of sight (Bar-Kana 1996, Keeton et al. 1997).

However, the use of elliptical mass distributions has not become common
practice because the evaluation of the deflection angle and
magnification matrix requires some effort. Bourassa et al.\ (1973)
and Bourassa and Kantowski (1975) (with minor corrections by Bray
(1984)) introduced a complex
formulation of lensing which allows for an elegant expression of
the deflection angle due to a homoeoidal elliptical mass 
distribution. I.e., this is a projected, two dimensional mass 
distribution whose isodensity contours are concentric ellipses of
constant ellipticity and orientation. It can be obtained, e.g.,
by projecting a three-dimensional homoeoidal mass distribution.
The complex integral which gives the deflection angle is in
practice difficult to separate into real and imaginary parts.
Schramm (1990) used an alternative derivation to obtain the
deflection without the use of complex numbers, but still
requiring a numerical integral for each component of the
deflection angle. Elliptical densities have been used for numerical
lens modelling, with the ellipticity allowed to vary in order
to fit the data, by Keeton \& Kochanek (1997).

In order to avoid numerical integration, several alternatives
have been suggested to exact elliptical mass distributions.
Models where the potential is chosen to have elliptical
contours rather than the density are easy to use, since
the deflection can be obtained immediately as the gradient
of the potential.
The imaging properties of elliptical potentials have been 
investigated extensively (Kovner 1987, Blandford \& Kochanek
1987 and Kochanek \& Blandford 1987). They become identical
to elliptical densities for very small ellpticities and
produce similar image configurations even for moderate
ellipticty (Kassiola \& Kovner 1993, hereafter KK93). However, elliptical
potentials cannot represent mass distributions with axis
ratios $b/a$ smaller than about $0.5$ because the 
corresponding density contours acquire the artifical 
feature of a dumbbell shape, and the density can also
become negative in some cases (Kochanek \& Blandford 1987,
KK93). To avoid this problem, Schneider
\& Weiss (1991) proposed a numerical method based on a
multipole expansion of the mass distribution, but this
expansion converges slowly when used with large ellipticity.

In this paper we consider a family of projected density 
profiles that has been used with the approximate 
approaches to ellipticity discussed above. This is the family 
of softened power-law profiles, which have a constant density
within a core radius and approach a power-law fall-off at
large radii. In \S 2 we introduce our notation for
softened power-law elliptical mass distributions (SPEMDs)
and for softened power-law elliptical potentials (SPEPs),
and illustrate further the limitations of SPEPs. In \S 3
we present and simplify the quadrature solution of Schramm (1990) 
for the deflection angle. We show that for the SPEMDs it
is possible to approximate the integrand so that the
integral can be done analytically. Although the result is
a sum of series expansions, each series converges
rapidly to high accuracy, even for mass densities with
arbitrarily high ellipticity. We show how to similarly
evaluate the magnification matrix. We also derive an
expression for the gravitational potential, but it
cannot be evaluated without a numerical integration.
Finally, in \S4 we summarize our results.

\section{SPEMDs and SPEPs}

Consider a projected mass density $\Sigma$ at an angular 
diameter distance $D_d$ and a source at $D_s$, with a
distance of $D_{ds}$ from the lens to the source. Then
the lens equation can be written as 
\beq
\by=\bx-\bal(\bx)
\eeq
where $\by$ is the source angle, $\bx=(x_1,x_2)$
is the observed image angle, and $\bal=(\al_1,\al_2)$ is the
deflection angle scaled by a factor of $D_{ds}/D_s$ (see
e.g.\ Schneider et al.\ 1992 for an introduction to gravitational
lensing).  The deflection is 
\beq
\bal(\bx)={\bf \nabla}\psi(\bx),
\eeq
in terms of the potential 
\beq
\psi(\bx)=\frac{1}{\pi}\int d^2x' \kappa({\bf x'})
\ln |\bx-{\bf x'}|\ .
\label{psi}
\eeq
Equation \ref{psi} is the solution to the Poisson equation
\beq \nabla^2\psi(\bx)=2 \kappa(\bx)\ , \label{pois}\eeq where 
$\kappa=\Sigma/\Sigma_{cr}$ in terms of the critical density
\beq \Sigma_{cr}=\frac{c^2}{4 \pi G}\frac{D_s}{D_d D_{ds}}\ .
\eeq

The SPEMD is given by \beq
\kappa(\rho)=\left[\frac{\rho^2+s^2}{E^2}\right]^{\frac{\eta}{2}-1}
\ , \eeq where $s$ is the core radius, $\eta$ is the power-law index, 
and $E$ fixes the overall normalization. The
dependence on position is through \beq \rho^2=x_1^2+x_2^2/\cos^2\beta
\ , \label{eqrho} \eeq where $\cos \beta$ is the axis ratio $b/a$ and 
the ellipticity is $e=1-\cos \beta$. The parameter $\eta$ can vary from a 
modified Hubble profile ($\eta=0$) through isothermal ($\eta=1$) to a 
constant surface density sheet ($\eta=2$). 

The SPEP is specified by a potential \beq \psi(\rho_p)=\frac{2
E_p^2}{\eta^2}\left[\frac{\rho_p^2+s_p^2}{E_p^2}\right]^
{\frac{\eta}{2}}\ ,\eeq where \beq \rho_p^2=x_1^2+x_2^2/\cos^2\beta_p
\ . \eeq The corresponding density as specified by the Poisson
equation (\ref{pois}) is \beq \kappa_p=\frac{1}{\eta}
\left[\frac{\rho_p^2+s_p^2}{E_p^2}\right]^{\frac{\eta}{2}-1}
\left\{(\eta-2)\frac{x_1^2+x_2^2/\cos^4\beta_p}{\rho_p^2+s_p^2}+
\left(1+\frac{1}{\cos^2\beta_p}\right)\right\}\ .\eeq
For $\rho_p \gg s_p$, the axis ratio of this density distribution
is \beq A_p=\cos \beta_p \left[\frac{1+(\eta-1)\cos^2 \beta_p}{\eta
-\sin^2\beta_p}\right]^{\frac{1}{\eta-2}} \eeq (Equivalent forms
for $\kappa_p$ and $A_p$ have been derived by KK93 and Grogin and Narayan
(1996), respectively). KK93 have
shown that the isodensity contours acquire a concave part
and become dumbbell shaped when $\cos^2 \beta_p < 1-\eta/3$. Thus
for a given $\eta$ the requirement of convex density contours
implies a smallest axis ratio $A_p$ that can be represented, or
equivalently a maximum ellipticity of \beq e_{\rm max}(\eta)=1-
\sqrt{1-\frac{\eta}{3}}\left(2-\frac{\eta}{2}\right)^{\frac{1}
{\eta-2}}\ . \eeq Figure 1 plots this maximum ellipticity 
as a function of $\eta$, and shows that $e_{\rm max}$ is typically
around $0.5$. An ellipticity greater than $0.65$ cannot be
modelled for any $\eta<2$. 

\begin{figure}[hbt]
\begin{center}
\centerline{\epsfxsize=11cm\epsfbox{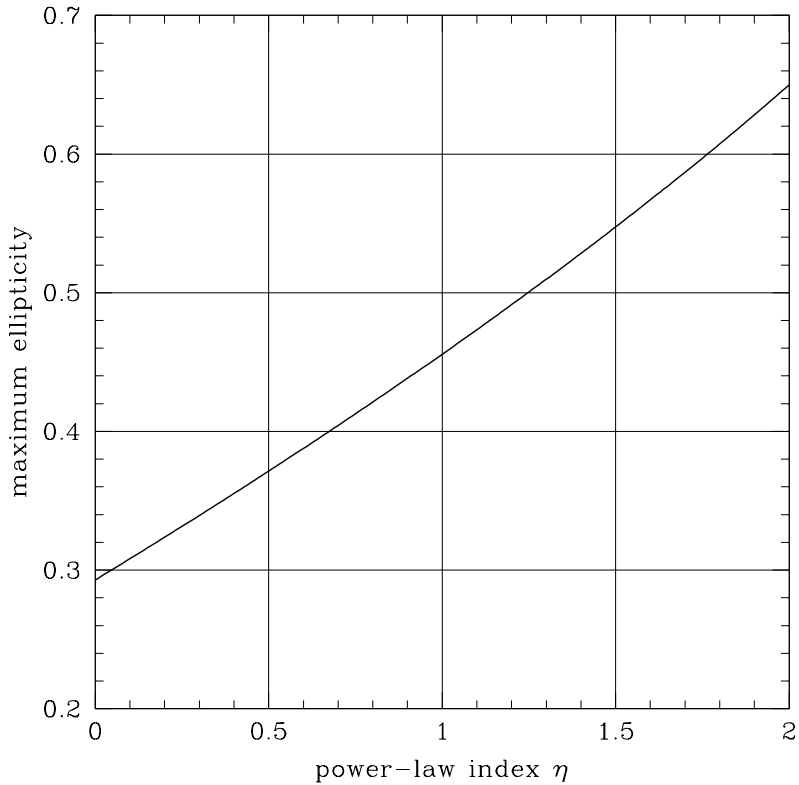}}
\caption{
Maximum ellipticity representable by an SPEP model shown as 
a function of the power-law index $\eta$. }
\end{center}
\label{femax}
\end{figure}

Assigning corresponding parameters between the SPEP and the SPEMD
is somewhat arbitrary, but we follow KK93 and use the following
conventions. For zero core radius and zero ellipticity, we
require the SPEP and SPEMD to match, which yields the condition
$E=E_p$. We require the density axis ratios $b/a$ far from the core
to be equal, i.e.\ $\cos \beta=A_p$. Finally, we also
require the central densities to match (if the core
radii are nonzero), which yields $$ s=s_p \left[\frac{1+
\cos^2\beta_p}{\eta \cos^2\beta_p}\right]^{\frac{1}{\eta
-2}}\ . $$ Figure 1 shows that at a fixed $\eta$,
as the ellipticity is increased the contours of the SPEP
become dumbbell shaped at the critical value $e_{\rm max}
(\eta)$. This is illustrated in Figure 4 of KK93, which shows
the density contours of several pairs of corresponding SPEMDs 
and SPEPs, for several values of $b/a$ and with $\eta$ fixed at 1.
Figure 2 illustrates the complementary case, where 
if we decrease $\eta$ at a fixed ellipticity the contours
switch over from being convex to having the dumbbell shape.
Figure 2a shows the isodensity contours of an SPEMD
with axis ratio $b/a=0.5$. For simplicity, we set the core radius
equal to zero. For the SPEMD, the contour shape
depends only on the axis ratio and is independent of $\eta$.
Figure 2b shows the isodensity contours of the 
corresponding SPEP with $\eta=1.6$ and the same axis ratio of 
$0.5$ for the density contours. As we decrease $\eta$
keeping the axis ratio fixed, the SPEP contours remain
convex until the value of $\eta=1.25$ shown in Figure 2c,
but they become dumbbell shaped below this as illustrated
in Figure 2d for $\eta=0.4$.

\begin{figure}[hbt]
\begin{center}
\centerline{\epsfxsize=11cm\epsfbox{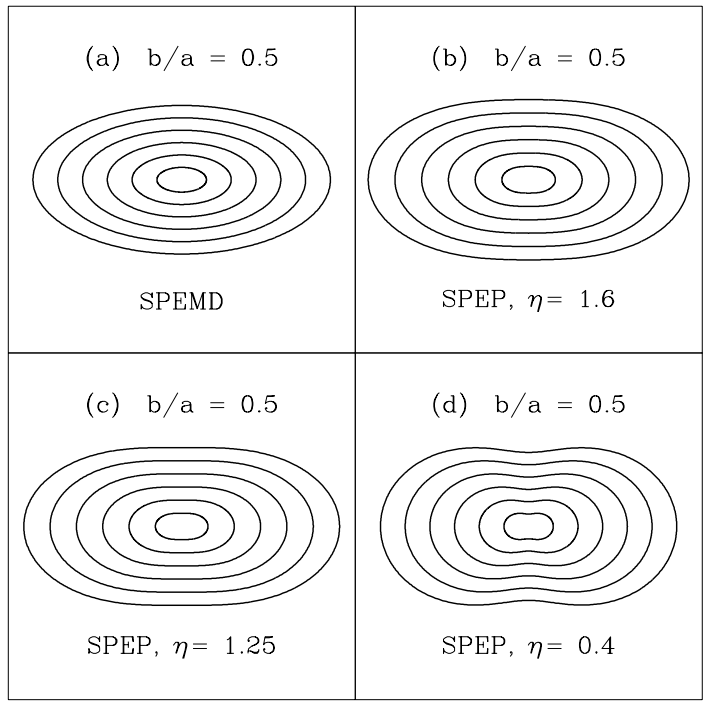}}
\caption{
Isodensity contours of SPEMDs and corresponding SPEPs. Panel (a) shows
the SPEMD contours with axis ratio $0.5$. For a fixed axis ratio,
as we decrease $\eta$ the SPEP contours switch from convex to dumbbell
shaped, as shown in panels (b), (c), and (d).}
\end{center}
\label{fcon}
\end{figure}

\begin{figure}[hbt]
\begin{center}
\centerline{\epsfxsize=11cm\epsfbox{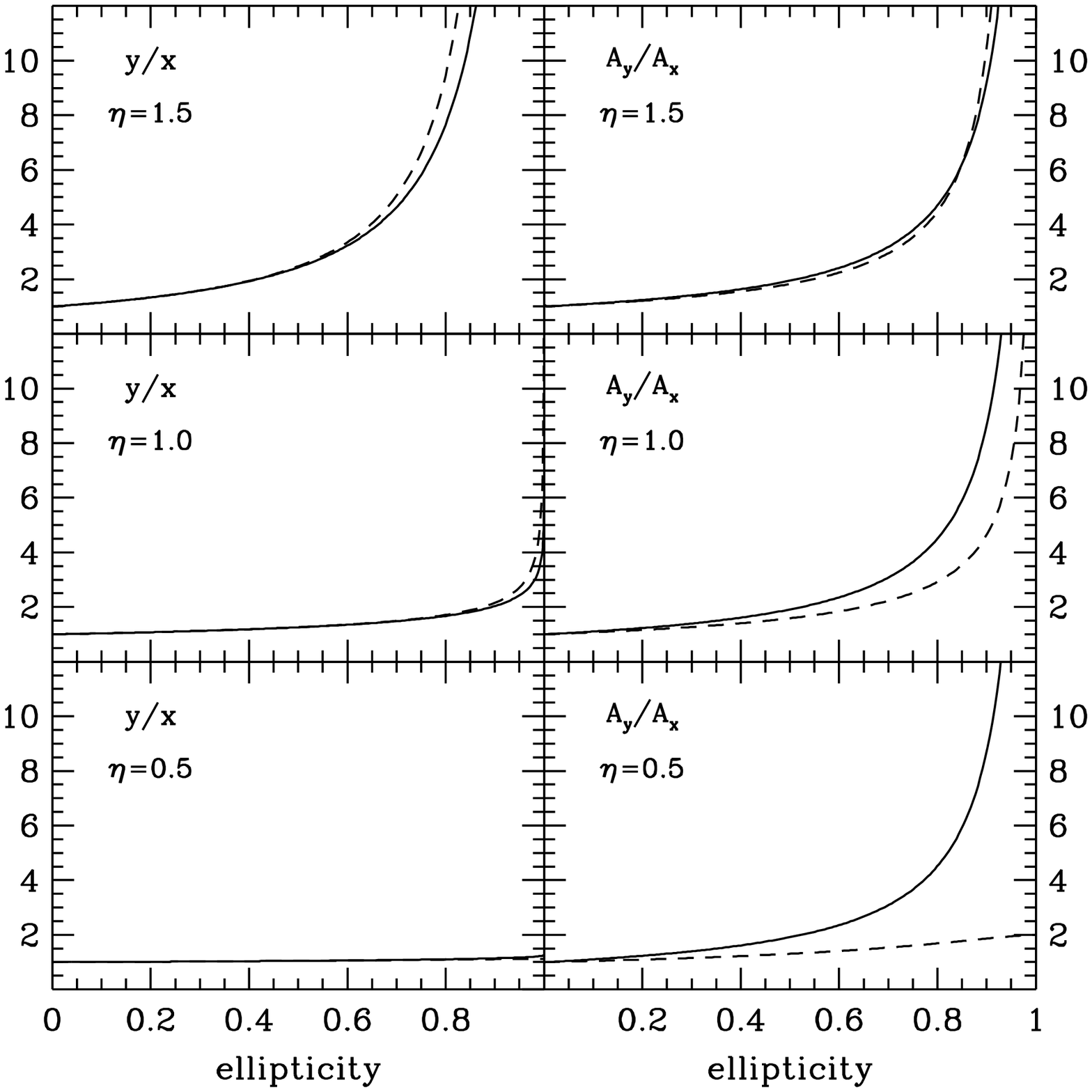}}
\caption{
Properties of the four images of a source placed directly
behind the lens center,
for SPEMDs and corresponding SPEPs. Several different power-law indices
$\eta$ are considered. The panels on the left show the 
distance $y$ of the images on the $y$-axis over the distance $x$ of 
the images on the $x$-axis, and the panels on the right show the
magnification ratio $A_y/A_x$ of the two types of images. In each
case the appropriate quantity is plotted as a function of ellipticity,
with the SPEMD as the solid curve and the SPEP as the dashed
curve.}
\end{center}
\label{fkk}
\end{figure}

Even with the problem of the SPEP regarding isodensity contour shape, 
it is of interest to directly compare the lensing behaviour of the
two models, the SPEP and SPEMD. We compare them in the same 
configuration as KK93, i.e.\ for a source placed directly behind
the center of the lens. In this case there are four images, with two
images on the y-axis at $(0,\pm y$) with magnification $A_y$
and two on the x-axis at $(\pm x,0)$ with magnification $A_x$.
Figure 7 of KK93 shows the distance ratio $x/y$ and the magnifications
$A_x$ and $A_y$ for the SPEMD and corresponding SPEP, as a function of
ellipticity for the singular (i.e.\ zero core) isothermal profile. In 
figure 3 we plot the distance ratio $y/x$ and the magnification 
ratio $A_y/A_x$ for singular profiles with $\eta=0.5$, $\eta=1$, 
and $\eta=1.5$. The distance ratios behave similarly for the SPEMD 
(solid curves) and SPEP (dashed curves), though there are small 
differences at high ellipticity.
The magnification ratio is an observable, not the individual 
image magnifications, and we find that the behavior of the 
magnification ratio at high ellipticity differs significantly between 
the SPEMD and SPEP even at $\eta=1$, but more so at smaller $\eta$. 
Thus while the SPEP produces some image configurations 
qualitatively similar to 
those of the SPEMD, quantitatively it is not an accurate substitute
unless the ellipticity is small.

\section{The deflection angle, magnification matrix, and potential
of the SPEMD}

We begin with the solution of Schramm (1990) for any elliptical
density of the form $\kappa(\rho)$, which we write concisely as 
\beqa \alpha_1(x_1,x_2)&=&2 x_1\,\cos\beta\, \int_0^{\rho(x_1,
x_2)}\frac{\rho '  \kappa(\rho ')\, \omega}{x_1^2+
\omega^4\, x_2^2 } d\rho ' \ , \nonumber \\ 
\alpha_2(x_1,x_2)&=&2 x_2\,\cos\beta\, \int_0^{\rho(x_1,x_2)}
\frac{\rho ' \kappa(\rho ')\, \omega^3}{x_1^2+
\omega^4\, x_2^2 } d\rho ' \ , \label{schr} \\
\omega^2 &\equiv& \frac{\Delta +r^2+\rho '^2\, \sin^2
\beta}{\Delta +r^2-\rho '^2\, \sin^2\beta}\ , 
\nonumber \\ \Delta^2 &\equiv& \left[\rho '^2\, \sin^2\beta
+x_2^2-x_1^2\right]^2+4 x_1^2 x_2^2\ , \nonumber \eeqa
where $r^2=x_1^2+x_2^2$.
The upper limit of integration $\rho(x_1,x_2)$ is given by 
equation (\ref{eqrho}).
In what follows we can restrict $x_1$ and $x_2$ to be nonnegative 
without loss of generality, noting from equations (\ref{schr}) that
$\al_1(x_1,x_2)=\al_1(|x_1|,|x_2|)\,\sign(x_1)$ and
$\al_2(x_1,x_2)=\al_2(|x_1|,|x_2|)\,\sign(x_2)$.

To simplify $\Delta$, we switch variables to \beq \label{mu} \mu=
\frac{\rho '^2\, \sin^2\beta+x_2^2-x_1^2}{2 x_1 x_2}\ . \eeq
We find that $\omega^2$ factorizes as $\omega^2=
(\mu+\sqrt{\mu^2+1})\, x1/x2$, which is a key point for our
method below to work. If we write $\kappa$ as a 
function of $\mu$, i.e.\ $\bar{\kappa}(\mu)\equiv\kappa(\rho)$,
we obtain
\beqa
\alpha_1(x_1,x_2)&=&\frac{\cos\beta\sqrt{x_1x_2}}{\sin^2\beta}
\int_{\mu_1}^{\mu_2}\bar{\kappa}(\mu)f(\mu)d\mu\ , \nonumber \\
\alpha_2(x_1,x_2)&=&\frac{\cos\beta\sqrt{x_1x_2}}{\sin^2\beta}
\int_{\mu_1}^{\mu_2}\bar{\kappa}(\mu)f(-\mu)d\mu\ , \nonumber \\
f(\mu) &\equiv& \sqrt{\frac{1}{\sqrt{1+\mu^2}}-\frac{\mu}{\mu^2
+1}}\ , \\
\mu_1 &\equiv& \frac{1}{2}\left(\frac{x_2}{x_1}-\frac{x_1}{x_2}
\right)\ , \nonumber \\
\mu_2 &\equiv& \frac{1}{2}\left(\frac{x_2}{x_1\cos^2\beta}-\frac
{x_1\cos^2\beta}{x_2}\right)\ . \nonumber \eeqa

This analysis so far is correct for any density of the form
$\kappa(\rho)$. We now specialize to the SPEMD, which we can
write as $\bar{\kappa}(\mu)=\bar{q}(\mu+\sb)^{-\gamma}$. In
terms of our variables from \S 2, we have $\gamma=1-\eta/2$, 
$\bar{q}=[2 x_1x_2/(E^2\sin^2\beta)]^{-\gamma}$, and 
$\sb=-\mu_1+s^2\sin^2\beta/(2 x_1x_2)$. Defining $\nu\equiv
\mu+\sb$, we finally obtain \beqa \alpha_1(x_1,x_2)&=&
\widetilde{q} I_1\ , \nonumber \\ \alpha_2(x_1,x_2)&=& 
\widetilde{q} I_2\ , \label{f1}\eeqa 
in terms of a common factor \beq \widetilde{q} = \frac{\cos\beta
\sqrt{x_1x_2}}{\sin^2\beta}\bar{q} \eeq and the integrals \beqa 
I_1&=&\int_{\nu_1}^{\nu_2} \nu^{-\gamma} f(\nu-\sb)d\nu\ , 
\nonumber \\ I_2&=&\int_{\nu_1}^{\nu_2} \nu^{-\gamma} f(\sb-\nu)
d\nu\ , \eeqa evaluated between the limits
\beq \nu_1=\frac{s^2\sin^2\beta}{2 x_1x_2}\ ;\hspace{.4in}
\nu_2=\nu_1+\frac{\sin^2\beta}{2}\left(\frac{x_1}{x_2}+
\frac{x_2}{x_1 \cos^2\beta}\right)\ . \label{f2} \eeq

To evaluate the integral $I_1$ we make use of the fact that
$f(\mu)$ depends only on the variable $\mu$ and we know
its functional form, so we can approximate
$f(\nu-\sb)$ as a polynomial in either $\nu$ or $1/\nu$
for various ranges of $\nu$ from $0$ to $\infty$ (the
lower limit $\nu_1$ in $I_1$ is always nonnegative).
After some experimentation, we settle on a number of
strategies for these polynomial expansions.

For a small range of $\mu$ values around $0$, e.g.\ $\mu_a$ 
to $\mu_b$, we use Chebyshev polynomials (see e.g.\ Press et al.\
1992) to construct a
polynomial approximation for $f(\mu)$ of the form $\sum_{n=0}^
{n=N} c_n \mu^n$. Then we can evaluate the portion of $I_1$ 
between $\nu_a=\mu_a+\sb$ and $\nu_b=\mu_b+\sb$ to be (e.g.\ 
if $\gamma$ is not an integer) $$ \int_{\nu_a}^{\nu_b} \nu^{-\gamma}
\sum_{n=0}^{n=N}c_n(\nu-\sb)^n d\nu=\sum_{n=0}^{n=N}\sum_{m=0}^
{m=n}\frac{c_n n!}{m!(n-m)!}\frac{(-\sb)^{n-m}}{m+1-\gamma}\left(
\nu_b^{m+1-\gamma}-\nu_a^{m+1-\gamma}\right)\ . $$ 
We can reduce the number of computations if $\sb \gg |\mu_b-\mu_a|$, 
because we can then write $\nu^{-\gamma}=\sb^{-\gamma}
[1+(\nu-\sb)/\sb]^{-\gamma}$, expand this in powers of $(\nu-\sb)/
\sb$, and
use $\nu-\sb$ as the variable of integration. The advantage of
using Chebyshev polynomials is that the 
expansion converges rapidly if the range of $\mu$ values is not 
too large.

For $\mu \gg 1$, we expand $f(\mu)$ in inverse powers of $\mu$,
\beq f(\mu)\approx\frac{1}{\sqrt{2}\mu^{3/2}}-\frac{5}{8\sqrt{2}
\mu^{7/2}}\ . \label{exp1}\eeq Consider just the first term (we deal 
similarly with the second term). We can write a contribution to $I_1$ in 
this regime as $$ \int_{\nu_a}^{\nu_b} \nu^{-\gamma} \frac{d\nu}{\sqrt{2}
(\nu-\sb)^{3/2}}=\frac{1}{\sqrt{2}}|\sb|^{-\gamma-1/2}\int_{\nu_a/|\sb|}
^{\nu_b/|\sb|} \nb^{-\gamma} \frac{d\nb}{(\nb-\ub)^{3/2}}\ , $$ where
we have set $\nb\equiv\nu/|\sb|$ and $\ub\equiv\sign(\sb)$ is $\pm 
1$. Then if $\nb\gg 1$, we expand $$(\nb-\ub)^{-3/2}=\nb^{-3/2} 
\left(1+\frac{3\ub}{2\nb}+\frac{15}{8\nb^2}+\ldots\right)$$ and then
integrate. On the other hand, for small values of $\nb$ up to a few we
can again use Chebyshev polynomial approximation, this time for the
function $(\nb-\ub)^{-3/2}$, and then integrate (we do this separately
for the two signs of $\ub$). In practice we find 
that it is better to perform an integration by parts and then use 
a polynomial approximation for the function $(\nb-\ub)^{-1/2}$
instead. When $\ub=+1$ we cannot extend this down to $\nb=1$, so for a
range of $\nb$ near 1 we leave $(\nb-1)^{-3/2}$ unchanged but
Taylor expand the $\nb^{-\gamma}$ term about $\nb=1$. The result
is an integrand which is a sum of powers of $\nb-1$ and is
easily integrated. When $\ub=-1$, we find that we sometimes
need high accuracy near $\nb=0$, so for a small range of
$\nb \ll 1$ we Taylor expand $(\nb+1)^{-3/2}$ about the
midpoint of this narrow range, and then integrate.

We use the same set of methods when $(-\mu) \gg 1$, but here the
expansion is \beq f(\mu)\approx\frac{\sqrt{2}}{\sqrt{-\mu}}
-\frac{3}{4\sqrt{2}(-\mu)^{5/2}}\ .\label{exp2}\eeq

In some cases we can use other methods to increase the speed.
If the entire integration range of $I_1$ is narrow, i.e.\ 
$(\nu_2-\nu_1) \ll 1$, then we Taylor expand $f(\nu-\sb)$
about the midpoint of the range and then integrate, without
breaking up this range into the various regimes described 
above.

The required accuracy determines how we specify the various
regimes. We were motivated by the lens 0957+561 (e.g.\ 
Walsh et al.\ 1979, Young et al.\ 1981, Falco et al. 1991, 
Grogin \& Narayan 1996), where VLBI observations (Porcas et al.\ 
1981, Gorenstein et al.\ 1988, Garret et al.\ 1994)
require a relative accuracy in the deflection angle of
one part in $10^5$. For algorithmic simplicity we prefer to
use only two terms in the expansions (\ref{exp1}) and
(\ref{exp2}), so this high accuracy forces us to use them
only for $|\mu| > 15.7$. For $-15.7 < \mu < 15.7$ we
use the direct Chebyshev expansion of $f(\mu)$. We find
it best to break up this range into 7 intervals and
construct a Chebyshev expansion in each, with the
number of terms in the expansions ranging from 6 to
9, keeping the relative accuracy roughly constant.

The integral $I_2$ is evaluated using exactly the same 
expansions as $I_1$ but with $f(-\mu)$ substituted for
$f(\mu)$. The components of the $2\times 2$ inverse magnification 
matrix $\delta_{ij}-\partial\al_i/\partial x_j$ also yield similar 
integrals. Using equations (\ref{f1}) through (\ref{f2}) we derive
\beqa \frac{\partial \al_1}{\partial x_1}&=&\left(\frac{1}
{2}-\gamma\right)\frac{\al_1}{x_1}+\widetilde{q}
\left[\nu_2^{-\gamma} f(\nu_2-\sb)\frac{\partial \nu_2}
{\partial{x_1}}-\nu_1^{-\gamma} f(\nu_1-\sb)\frac{\partial 
\nu_1} {\partial{x_1}}-I_3 \frac{\partial \sb}{\partial 
x_1}\right]\ , \nonumber \\
\frac{\partial \al_2}{\partial x_2}&=&\left(\frac{1}
{2}-\gamma\right)\frac{\al_2}{x_2}+\widetilde{q}
\left[\nu_2^{-\gamma} f(\sb-\nu_2)\frac{\partial \nu_2}
{\partial{x_2}}-\nu_1^{-\gamma} f(\sb-\nu_1)\frac{\partial 
\nu_1} {\partial{x_2}}+I_4 \frac{\partial \sb}{\partial 
x_2}\right]\ , \label{magmx}
\\ \frac{\partial \al_2}{\partial x_1}&=&
\frac{\partial \al_1}{\partial x_2}=\left(\frac{1}
{2}-\gamma\right)\frac{\al_1}{x_2}+\widetilde{q}
\left[\nu_2^{-\gamma} f(\nu_2-\sb)\frac{\partial \nu_2}
{\partial{x_2}}-\nu_1^{-\gamma} f(\nu_1-\sb)\frac{\partial 
\nu_1} {\partial{x_2}}-I_3 \frac{\partial \sb}{\partial 
x_2}\right]\ . \nonumber \eeqa Just as with $I_1$ and
$I_2$, we evaluate the integrals 
\beqa I_3&=&\int_{\nu_1}^{\nu_2}\nu^{-\gamma}f '(\nu-\sb)
d\nu \ , \nonumber \\ I_4&=&\int_{\nu_1}^{\nu_2}\nu^{-\gamma}
f '(\sb-\nu)d\nu \ , \eeqa where we approximate $f '(\mu)\equiv
d\/f(\mu)/d\mu$ in the same ways as we created polynomial
approximations for $f(\mu)$ above. The other quantities
appearing in equations (\ref{magmx}) can be evaluated 
(for $i=1$ and 2) as
\beqa \frac{\partial \nu_1}{\partial x_i}&=&-\frac{\nu_1}
{x_i}\ ;\hspace{.4in} \frac{\partial \sb}{\partial x_i}=
\frac{\partial \nu_1}{\partial x_i}-\frac{1}{2 x_i}
\left[\frac{x_1}{x_2}+\frac{x_2}{x_1}\right](-1)^i
\nonumber \\ \frac{\partial \nu_2}{\partial x_i}&=&
\frac{\partial \nu_1}{\partial x_i}-\frac{\sin^2\beta}
{2 x_i}\left[\frac{x_1}{x_2}-\frac{x_2}{x_1\cos^2\beta}
\right](-1)^i\ . \eeqa 

If we apply the same amount of computational effort to
$I_3$ and $I_4$ as for $I_1$ and $I_2$, we 
nevertheless find that the relative accuracy is lower
for the magnification det$^{-1}|\delta_{ij}-\partial\al_i/
\partial x_j|$
than it is for the components of $\bal$, for a
couple reasons. First, the various polynomial expansions
tend to converge more slowly for $f '(\mu)$ than
for $f(\mu)$; and second, the components of $\bal$
are directly proportional to $I_1$ and $I_2$, but in 
equations (\ref{magmx}) we sometimes have to subtract nearly
equal terms and get a low accuracy for $\partial\al_i/
\partial x_j$
even if $I_3$ and $I_4$ are evaluated accurately.
In practice we achieve a maximum relative error 
of $5\times 10^{-6}$ for $\bal$ and $6\times
10^{-4}$ for the magnification, and a typical 
relative error of $1\times 10^{-6}$ for $\bal$ and 
$5\times10^{-5}$ for the magnification. A higher 
accuracy is not needed for the magnification, given 
typical measurement errors on fluxes. 

In terms of running time, with the above accuracy we 
can evaluate the deflection angle and magnification 
matrix of the SPEMD roughly 20 times faster than the 
brute force method (which requires 5 integrals), 
although this is still about
15 times slower than the SPEP, or any other
model with the deflection given by a simple formula.
This speedup should make the SPEMD model useful for applications
in which repeated numerical integrations made it
previously unusable.

To evaluate the gravitational part of the time delay, we also
need to evaluate the potential $\psi(\bx)$. Schramm (1990) gives
a quadrature expression for the potential of a single elliptical 
shell. We obtain the potential of any elliptical density 
$\kappa(\rho)$ by integrating over shells,
\beq \psi(x_1,x_2)=\cos \beta \int_{\rho ' = 0}^{\rho(x_1,x_2)}
\int_{u=0}^{\gamma(\rho ')} \frac{d u}{\sqrt{(\rho '^2+u)(\rho '^2
\cos^2\beta+u)}}\rho ' \kappa(\rho ') d\rho '\ , \eeq
where \beq \gamma(\rho ')=\frac{1}{2}\left\{r^2-\rho '^2
(1+\cos^2\beta)+\Delta\right\} \eeq in terms of the last equation 
of (\ref{schr}). Performing the inner integral, we derive
\beq \psi(x_1,x_2)=2 \cos \beta \int_{\rho ' = 0}^{\rho(x_1,x_2)}
\ln\left[ \frac{\sqrt{\Delta+r^2-\rho '^2\sin^2\beta}+
\sqrt{\Delta+r^2+\rho '^2\sin^2\beta}}{\sqrt{2}\rho '
(1+\cos\beta)}\right]\rho ' \kappa(\rho ') d\rho ' \ . \eeq
In this case the integrand does not factorize when we
substitute equation (\ref{mu}), so the potential must be
numerically integrated and cannot be speeded up. In most
applications, though, it is not necessary to evaluate the
potential a large number of times.

\section{Conclusions}

A mass density profile with elliptical isodensity contours is 
a natural lens model to try when axisymmetric models fail.
Previously, however, the easiest way of evaluating the 
deflection has been to numerically integrate the 
solutions of Schramm (1990). After simplifying these 
solutions we have shown that for the family of
SPEMD mass distributions, the deflection angle and
magnification matrix can be evaluated very fast
and with high accuracy. Our implementation 
achieves a relative accuracy of $5\times 10^{-6}$
in the deflection and $6\times 10^{-4}$ in the magnification
while running 20 times faster than a procedure based on
the numerical integrations. We have also derived an
expression for the potential, although this quantity must 
be numerically integrated.
 
As noted by Schneider
\& Weiss (1991), combinations of two or more SPEMDs 
with different parameters can be used to construct more 
general density profiles with several scales. 
We thus expect SPEMDs to be more widely used, 
particularly for cases of high ellipticity in which
the alternative SPEPs develop the artificial feature
of dumbbell-shaped contours.

\acknowledgements

I thank Ed Turner for valuable discussions.
This work was supported by Institute Funds.

\end{document}